\documentclass[onecolumn,amsmath,amssymb,nofootinbib,12pt]{article}
\usepackage{jheppub}
\usepackage{ifpdf}
\usepackage{graphicx}
\usepackage{amsfonts}
\usepackage{amsmath}
\usepackage{amssymb}
\usepackage{epsfig}
\usepackage{pdfpages}
\usepackage{graphicx,epstopdf}
\usepackage{caption}
\usepackage[position=b]{subcaption}
\usepackage[makeroom]{cancel}
\usepackage{hyperref}
\hypersetup{pdftex,colorlinks=true,allcolors=blue}
\usepackage{array}

\def\hor{\big|_{\mathcal{N}}}
\def\onb{\big|_{\mathcal{B}}}
\def\csform{\Theta}
\newcommand{\cM}{{\mathcal{M}}}
\newcommand{\cQ}{{\mathcal{Q}}}

\newcommand{\wh}[1]{\widehat{#1}}
\def\IZ {\mathbb{Z}}

\begin{document}
	
\preprint{arXiv:1803.04517 [hep-th]}
\title{T-duality and high-derivative gravity theories: the BTZ black hole/string paradigm}

\author[a]{Jos\'e D. Edelstein,}
\author[b]{Konstantinos Sfetsos,}
\author[c]{J. Anibal Sierra-Garcia,}
\author[a]{Alejandro Vilar L\'opez}
\affiliation[a]{Departamento de F\'\i sica de Part\'\i culas $\&$ Instituto Galego de F\'\i sica das Altas Enerx\'\i as (IGFAE), Universidad de Santiago de Compostela, E-15782 Santiago de Compostela, Spain}
\affiliation[b]{Department of Nuclear and Particle Physics, Faculty of Physics, National and Kapodistrian University of Athens, Athens 15784, Greece}
\affiliation[c]{Department of Physics, Faculty of Science, Chulalongkorn University, Bangkok 10330, Thailand}

%\emailAdd{jose.edelstein@usc.es}
%\emailAdd{ksfetsos@phys.uoa.gr}
%\emailAdd{anibal.sierra.garcia@gmail.com}
%\emailAdd{alejandro.vilar.lopez@gmail.com}
%\date{\today}

\abstract{We show that the temperature and entropy of a BTZ black hole are invariant under T-duality to next to leading order in $M_\star^{-2}$, $M_\star$ being the scale suppressing higher-curvature/derivative terms in the Lagrangian. We work in the framework of a two-parameter family of theories exhibiting T-duality, which includes (but goes beyond) String Theory. Interestingly enough, the AdS/CFT correspondence enforces quantization conditions on these parameters. In the particular case of bosonic/heterotic string theory, our results extend those of a classical paper by Horowitz and Welch. For generic (albeit quantized) values of the parameters, it suggests that T-duality might be an interesting tool to constrain consistent low-energy effective actions while entailing physical equivalences outside String Theory. Moreover, it generates a new family of regular asymptotically flat black string solutions in three-dimensions.}

\maketitle
%%%%%%%%%%%%%%%%%%%%%%
\section{Introduction}
%%%%%%%%%%%%%%%%%%%%%%

Despite its remarkable success, General Relativity is not amenable to the laws of Quantum Mechanics and there is a general understanding that it is the low energy description of another theory. As such, it is expected to receive higher-derivative corrections that are suppressed by some high-energy scale, $M_\star$. If $M_\star$ is well below the Planck mass and gravity remains weakly coupled, it was recently shown that both quadratic and cubic higher (Riemann) curvature corrections entail causality violation unless an infinite tower of higher-spin particles is introduced \cite{CEMZ}. An example is given by perturbative String Theory \cite{DAppollonio}. In this context, it is interesting to explore whether higher curvature Lagrangians that are not strictly in the realm of String Theory, but possess some of its distinctive features, may provide consistent descriptions of the low energy dynamics of a gravitational system. In this letter we will focus on T-duality.

T-duality is an equivalence between two string theories in different backgrounds. In particular, it relates string theory dynamics in geometries with large and small compact directions by identification of momentum and winding modes. Its uses range from the attempts to resolve the cosmological singularity in early universe models \cite{Brandenberger:1988aj,Veneziano:1991ek,Gasperini:1992em} to modern applications in the framework of the AdS/CFT correspondence (see, for example, \cite{AldayMaldacena}) mostly due to its remarkable potential as a highly non-trivial solution generating technique. The convenience of writing down supergravity actions in a manifestly T-dual invariant way propelled the development of so-called Double Field Theory \cite{Hull:2009zb, Hull:2009mi, Hohm:2010jy, Hohm:2010pp} (see \cite{Geissbuhler:2013uka,Aldazabal:2013sca} for comprehensive reviews), a convenient scheme to, so to say, treat momentum and winding modes on an equal footing. In this paper we shall study a two-parameter family of four-derivative T-dual invariant Lagrangians that was derived in this context \cite{MarquesNunez}, building up in an earlier formulation developed by Hohm and Zwiebach \cite{Hohm2014}. T-duality constraints on the construction of higher derivative theories were further explored by these authors in \cite{{Hohm2015}}.

In String Theory, T-duality is a symmetry of the low energy effective action to all orders in $\alpha^\prime$. Consider, for instance, the action that governs the universal massless NS-NS sector,
\begin{equation}
\mathcal{I}_0 = \int d^D\!x \sqrt{-\mathrm{G}}\, e^{- 2 \Phi} \mathcal{L}^{(0)} ~,
\label{ActionNSNS}
\end{equation}
where we have set the Newton constant $16\pi G_{\rm N} = 1$, and $\mathcal{L}^{(0)}$ reads
\begin{equation}
\mathcal{L}^{(0)} = R - 2 \Lambda + 4 ( \nabla_M \nabla^M \Phi- \nabla_M \Phi \nabla^M \Phi ) - \frac{1}{12} H_{MNR}\,H^{MNR} ~.
\label{L0}
\end{equation}
The action involves the metric $\mathrm{G}_{MN}$, the dilaton, $\Phi$, and the Kalb-Ramond two-form potential, $B_{MN}$ (through its curvature, $H_{MNR} = \partial_M\,B_{NR} + \partial_N\, B_{RM} + \partial_R\, B_{MN}$). Besides, we have included a cosmological term which is T-dual invariant on its own and might be seen to arise from string compactification.\footnote{Our Riemann and Ricci tensor definitions are as in \cite{MarquesNunez}. $M, N, R, \ldots$ and $A, B, C, \ldots$ are $D$ dimensional (respectively curved and flat) $D$-dimensional indices, while lower-case indices $\mu,\nu,\rho,...$ and $a,b,c,...$ are $(D-1)$-dimensional.}

If we restrict our fields to possess a $\mathrm{U(1)}$ isometry along, say, $\psi$, there is a set of transformations given by so-called Buscher rules \cite{Buscher:1987sk} which leave (\ref{ActionNSNS}) invariant,
\begin{eqnarray}
& & \wh{\mathrm{G}}_{\psi\psi} = 1/\mathrm{G}_{\psi\psi} ~, \quad \wh{\mathrm{G}}_{\psi\mu} = B_{\psi\mu}/\mathrm{G}_{\psi\psi} ~, \quad \wh{B}_{\psi\mu} = \mathrm{G}_{\psi \mu}/\mathrm{G}_{\psi\psi} ~, \nonumber \\ [0.55em]
& & \wh{\mathrm{G}}_{\mu\nu} = \mathrm{G}_{\mu\nu} - (\mathrm{G}_{\psi\mu} \mathrm{G}_{\psi\nu} - B_{\psi\mu} B_{\psi \nu})/\mathrm{G}_{\psi\psi} ~, \\ [0.4em]
& & \wh{B}_{\mu\nu} = B_{\mu\nu} - (\mathrm{G}_{\psi\mu} B_{\psi \nu} - \mathrm{G}_{\psi\nu} B_{\psi \mu})/\mathrm{G}_{\psi\psi} ~, \quad \wh{\Phi} = \Phi - \frac{1}{2}  \log \mathrm{G}_{\psi\psi} ~, \nonumber
\label{BuscherRules}
\end{eqnarray}
where the hatted fields are those of the T-dual background. A more transparent expression for these rules emerges from the explicit writing of the fields in terms of those dimensionally reduced along $\psi$,
\begin{eqnarray}
& & d s^2 = g_{\mu\nu} dx^{\mu} dx^{\nu} + e^{2\sigma} (d\psi + V_\mu\,dx^\mu)^2 ~, \nonumber \\ [0.5em]
& & B_{\mu\nu} =  \mathcal{B}_{\mu\nu} + \frac{1}{2} \left( W_\mu V_\nu - W_\nu V_\mu \right) ~, \qquad B_{\mu\psi} = W_\mu ~, \label{KMansatz} \\ [0.5em]
& & \Phi = \phi + \frac{1}{2} \sigma ~, \nonumber
\end{eqnarray}
where $V_\mu$ and $W_\mu$ are the $U(1)$ gauge vectors arising from the metric and $B$-field components with an index along the cyclic coordinate \cite{KaloperMeissner}. $\mathcal{B}_{\mu\nu}$ is the dimensionally reduced Kalb-Ramond field, and the $(W_\mu V_\nu - W_\nu V_\mu)$-term is a convenient field redefinition \cite{MaharanaSchwarz} enforcing a manifestly simple set of T-duality rules,
\begin{equation}
V_\mu \leftrightarrow W_\mu ~, \qquad {\rm and} \qquad \sigma \leftrightarrow - \sigma ~,
\label{Tdualityrules}
\end{equation}
which make it transparent that the transformation squares to the identity.

One may ask whether T-duality provides a sufficient constraint to uniquely obtain the next-to-leading term in a low-energy expansion. In spite of our original expectations, this seems not to be the case as we will see below. Restricted to first-order corrections, a two parameter family of theories has been recently constructed by Marqu\'es and N\'u\~nez \cite{MarquesNunez} based on the $O(D,D)$ invariant Double Field Theory realization of T-duality,
\begin{equation}
\mathcal{L}^{(1)} = \gamma_+\,\mathcal{L}_+^{(1)} + \gamma_-\,\mathcal{L}_-^{(1)}  ~,
\label{TwoParameterLagrangian} 
\end{equation}
where $\gamma_\pm$ are assumed to be order $M_\star^{-2}$ parameters, and\footnote{$\mathcal{L}_+^{(1)}$ was first presented in \cite{MetsaevTseytlin} modulo some field redefinitions, integration by parts and neglection of boundary terms.}
\begin{eqnarray} \label{Loneplus}
\mathcal{L}_+^{(1)} &=& \frac{1}{2} R_{MNRS}\, R^{MNRS} - \frac{3}{4}  H^{MNR}\, H_{MSL}\, R_{NR}{}^{SL} + \frac{1}{6} \nabla_{M}\, H_{NRS}\,\nabla^M H^{NRS} \nonumber \\ [0.5em]
&& + \frac{1}{48} H^{MNR}\, H_{MS}{}^L\, H_{NL}{}^T\,H_{RT}{}^S + \frac{1}{16} H_{MRT}\,H^{MR}{}_L\,H_{NS}{}^T\,H^{NSL} \\ [0.5em]
&& - \frac{1}{2} H^{MNR} \,\partial_M \left( H_{NAB}\,\Omega_R^{\;\;\,AB} \right) ~,
 \nonumber \\ [0.7em] 
\label{Loneminus}\mathcal{L}_-^{(1)} &=& H^{MNR} \csform_{MNR} ~.
\end{eqnarray}
Notice that the sign subindex in $\mathcal{L}_\pm^{(1)}$ makes reference to its parity properties under the sign flip $H \to -H$. $\csform_{MNR}$ is the gravitational Chern-Simons three-form,
\begin{equation}
\csform_{MNR} = \Omega_{[MA}{}^B\,\partial_N\,\Omega_{R]B}{}^A + \frac{2}{3} \Omega_{[MA}{}^B\,\Omega_{NB}{}^C\,\Omega_{R]C}{}^A ~,
\label{CS3form}
\end{equation}
$\Omega_{MA}{}^B$ being the Lorentz connection, and the antisymmetrization is normalized, for example: $T_{[M N]}=\frac{1}{2} (T_{M N }-T_{N M})$. Some particular values of $\gamma_\pm$ correspond to low-energy effective actions coming from string theories:
\begin{eqnarray}
& & \gamma_- = 0 \quad\;\quad\,\qquad {\rm bosonic} ~, \nonumber \\ [0.3em]
& & \gamma_+ = - \gamma_- \quad\qquad {\rm heterotic} ~, \label{stringvalues} \\ [0.3em]
& & \gamma_+ = \gamma_- = 0 \qquad {\rm type~II} ~. \nonumber
\end{eqnarray}
In the former two cases, $\gamma_+ - \gamma_- = \frac{1}{2} \alpha^\prime$. The case $\gamma_+ = \gamma_-$ is also special \cite{HSZ}. We stress on the fact that for generic $\gamma_+$ and $\gamma_-$, not included in the previous cases, the Lagrangian above is not known to be related to a sigma model or a conformal field theory. Yet, it is invariant under T-duality if we neglect quadratic terms in $\gamma_\pm$. We will comment on the particular form of the T-duality rules below. At this stage, it is sufficient to mention that they also receive $\gamma_\pm$ corrections.

In this paper we would like to explore the behavior of the $\gamma_\pm$-corrected Ba\~nados-Teitelboim-Zanelli (BTZ) black hole \cite{BTZ}, which is a solution of the three-dimensional action
\begin{equation}
\mathcal{I} = \int d^3x \sqrt{-\mathrm{G}}\, e^{- 2 \Phi} \left[ \mathcal{L}^{(0)} + \gamma_+\,\mathcal{L}_+^{(1)} + \gamma_-\,\mathcal{L}_-^{(1)} \right] ~,
\label{Action}
\end{equation}
when it is subjected to a T-duality transformation. Horowitz and Welch \cite{HorowitzWelch-di} studied this problem when the dynamics is governed by (\ref{ActionNSNS}) or, in other words, when $\gamma_\pm = 0$. They found that the BTZ black hole is mapped onto a black string. Despite the geometry being severely modified, including its asymptotic behavior, the existence of a bifurcate Killing horizon, the temperature and the entropy of both solutions are the same. We want to scrutinize whether the result holds true when $\gamma_\pm$ corrections are included. It seems clear in \cite{HorowitzWelch-di} that the entropy being given by the area of the black hole horizon, thereby purely geometric in the Einstein frame, is at the basis of its invariance. We will show that this is not the case. The temperature and entropy of the BTZ black hole coincide with those of the T-dual black string when $\gamma_\pm$ corrections are included, the value of these parameters not being restricted to the string theory values (\ref{stringvalues}). Although T-duality does not provide stringent enough constraints on these parameters, the AdS/CFT correspondence does.

%%%%%%%%%%%%%%%%%%%%%%
\section{The BTZ black hole and $\gamma_\pm$ corrections}
%%%%%%%%%%%%%%%%%%%%%%

In order to discuss the effect of $\gamma_\pm$ corrections on the BTZ black hole, it is convenient to write down the equations of motion starting from an equivalent form of the action \eqref{Action} that is based on the construction of \cite{BergshoeffdeRoo},
\begin{eqnarray}
\mathcal{I} &=& \int d^3x \sqrt{-\mathrm{G}}\, e^{- 2 \Phi} \bigg[ R + \frac{4}{\ell^2} + 4 ( \nabla_M \nabla^M \Phi- \nabla_M \Phi \nabla^M \Phi ) - \frac{1}{12} \bar H_{MNR}\,\bar H^{MNR} \nonumber \\ [0.5em]
& & + \frac{1}{8} \sum_{k=\pm} a_k\, R^{(k)}_{M N A}{}^B R^{(k) M N}{}_B{}^A \bigg]  ~, \label{Lalphaprime}
\end{eqnarray}
where the sum runs over two signs, $k = \pm$, and the parameters $a_\pm$ are going to be dubbed $a_-  \equiv a$ and $a_+ \equiv b$.
\begin{equation}
\bar H_{MNR} := H_{MNR} - \frac{3}{2} \left( a\, \csform^{(-)}_{MNR} - b\, \csform^{(+)}_{MNR} \right) ~,
\label{Htilde}
\end{equation}
$\csform^{(\pm)}$ denoting the gravitational Chern-Simons form (\ref{CS3form}) of the Lorentz connection with torsion $\Omega^{(\pm)}_{MA}{}^B = \Omega_{MA}{}^B \pm \frac{1}{2} H_{M A}{}^B$. The precise definitions of $R^{(\pm)}_{M N A B}$ and $\csform^{(\pm)}_{M N R}$ are those of \cite{MarquesNunez}. The parameters $\gamma_\pm$ are related to $a$ and $b$ in a simple fashion:
\begin{equation}
\gamma_\pm = \mp \frac{a \pm b}{4} ~.
\end{equation}
Notice that the price to pay for writing the action as in (\ref{Lalphaprime}) is that it contains terms quadratic in $\gamma_\pm$; they must not be taken into consideration. From the variation of (\ref{Lalphaprime}) one gets the following equations of motion:
\begin{eqnarray}
& & R +\frac{4}{\ell^2}+ 4 ( \nabla^2 \Phi - (\nabla\Phi)^2 ) - \frac{1}{12} \bar H_{MNR}\,\bar H^{MNR} + \frac{1}{8} \sum_{k=\pm} a_k R^{(k)}_{M N A}{}^B R^{(k) MN}{}_B{}^A = 0 ~, \nonumber \\ [0.4em]
& & \nabla_{M} \left[  e^{-2\Phi}  \bar H^{MNR} + \frac{3}{2} \sum_{k=\pm} a_k \left( e^{-2\Phi} H^{ST[M}  R^{(k) RN]}_{ST} - k \nabla^{(k)}_{S} \left[ e^{-2\Phi} {R^{(k) S[MNR]}} \right] \right) \right] = 0 ~, \nonumber \\ [0.4em]
& & R_{MN} + 2 \nabla_M \nabla_N \Phi - \frac{1}{4} \bar H_{MRS} {\bar H_{N}}^{~RS} - \frac{1}{4} \sum_{k=\pm} a_k \left[ R^{(k)}_{MRST} R^{(k) RST}_N \right. \\ [0.4em]
& & \qquad\qquad \left. + e^{2\Phi} \left( 2 \mathrm{G}_{S(M|} \nabla_R + k H_{RS(M|} \right) \left( {\delta_U}^{S} \nabla^{(k)}_T + k {H_{TU}}^{S} \right) \left( e^{-2\Phi} {R^{(k) TUR}}_{|N)} \right) \right] = 0 ~, \nonumber
\label{full-eom}
\end{eqnarray}
where $\nabla^2 = \nabla_M \nabla^M$, $ (\nabla\Phi)^2 =  \nabla_M \Phi \nabla^M \Phi$, and $\nabla^{(k)}$ is the covariant derivative involving the connection with torsion\footnote{In other words, $\nabla^{(\pm)}$ is the (diffeomorphism) covariant derivative with torsionful Christoffel symbols $\Gamma_{M N}^{(\pm)R}=\Gamma_{M N}^R \mp \frac 1 2 H_{M N}{}^R$.} $\Omega^{(k)}_{MA}{}^B$. To begin with, we want to find exact black hole solutions of this system. The problem turns out to be straightforward if we start by recalling that the BTZ black hole \cite{BTZ} can be supplemented by a trivial dilaton and a quadratic Kalb-Ramond field to give a solution of the equations of motion arising from (\ref{L0}) \cite{HorowitzWelch-bh}.
\begin{eqnarray}
& & ds^2 = - N^2 dt^2 + \frac{dr^2}{N^2} + r^2 \left( d\psi + N^\psi dt \right)^2 ~, \nonumber \\ [0.4em]
& & e^{-2\Phi} = 1 ~, \label{BTZsolution} \\ [0.4em]
& & B_{t\psi} = \frac{r_+^2-r^2}{\ell} ~, \nonumber
\end{eqnarray}
while all other components vanish; the lapse is given by the standard BTZ form,
\begin{equation}
N^2 = \frac{(r^2-r_+^2)(r^2-r_-^2)}{\ell^2 r^2} ~,
\label{lapseBTZ}
\end{equation}
while
\begin{equation}
N^\psi = \frac{r_+ r_-}{\ell} \left( \frac{1}{r_+^2} - \frac{1}{r^2} \right) ~,
\label{Npsi}
\end{equation}
where $r_+$ and $r_-$ are two integration constants, better expressed in terms of the ($\gamma_\pm$-uncorrected) black hole mass and angular momentum computed from (\ref{ActionNSNS}),
\begin{equation} 
M = \frac{r_+^2 + r_-^2}{\ell^2} ~, \qquad J = \frac{2 r_+ r_-}{\ell} ~,
\label{bhMJ}
\end{equation}
which agree with the standard result \cite{BTZ}. Indeed, for the values in (\ref{stringvalues}), this solution is exact to all orders in $\alpha^\prime$ \cite{HorneHorowitzSteif}. The very fact that $a_\pm$ do not appear in (\ref{BTZsolution}), despite $\gamma_\pm \neq 0$ in bosonic/heterotic string theory, suggests that it can possibly be a solution for arbitrary $\gamma_\pm$. Indeed, this is the case. This happens because $\csform^{(\pm)}_{MNR}$ and $R^{(\pm)}_{\, MNAB}$ vanish when evaluated on (\ref{BTZsolution}), as expected. Notice that $H \to -H$ corresponds to a flip in the Lorentz connection with torsion $\Omega^{(\pm)}_{MA}{}^B \to \Omega^{(\mp)}_{MA}{}^B$.

In spite of the fact that the action (\ref{Lalphaprime}) can be shown to be local Lorentz invariant on-shell, it is not so for an arbitrary background. Indeed, it is invariant under an {\it anomalous} local Lorentz transformation of the form:
\begin{equation}
\delta_\Lambda E_M{}^A = E_M{}^B \Lambda_B{}^A ~, \qquad \delta_\Lambda B_{MN} = -\frac{a}{2}   \,\partial_{[M} \Lambda_A{}^B \Omega^{(-)}_{N] B}{}^A + \frac{b}{2} \,\partial_{[M} \Lambda_A{}^B \Omega^{(+)}_{N] B}{}^A ~,
\label{anomalouslorentz}
\end{equation}
where $\Lambda_A{}^B$ is the infinitesimal generator. For the heterotic string case (\ref{stringvalues}), $b = 0$ and $a = - \alpha^\prime$, such transformation is due to the Green-Schwarz anomaly cancellation mechanism. For the rest of the values of $a$ and $b$, it is formally similar.\footnote{In the case of bosonic string theory the necessity of an anomalous Lorentz transformation can be eliminated by means of a field redefinition. For more details, see \cite{MarquesNunez}.} Transformation (\ref{anomalouslorentz}) leaves $\bar H_{MNR}$ invariant, while the rest of the terms in the action are manifestly invariant. Due to this anomalous Lorentz invariance, it is necessary to specify both the vierbein\footnote{Besides the technical necessity of introducing a frame, it was recently shown that background independence plus the requirement of a manifest T-duality invariance call for its use \cite{Hohm2016a,Hohm2016b}.} and the $B$-field which solve the equations of motion. A different frame has to be accompanied by a compensating transformation in the $B$-field that in general changes its field strength $H_{M N R}$ so that $\bar{H}_{M N R}$ remains invariant. We shall see next that a $\Delta B_{t\psi}$, together with the choice of an appropriate reference frame, is necessary to bring our solution (\ref{BTZsolution}) to a regular form.

Consistency of the black hole thermodynamics requires the horizon to be a regular bifurcate Killing horizon \cite{RaczWald}. We consider a stationary solution with a spacelike isometry given by the Killing field $\eta$. The orbits of $\eta$ are closed and provide the symmetry along which we apply T-duality. Furthermore, $\eta$ commutes with an asymptotically timelike Killing field $\lambda$, and therefore we can associate coordinates to them as:
\begin{equation}
\lambda = \partial_t ~, \qquad \eta = \partial_{\psi'} ~.
\end{equation}
By definition of bifurcate Killing horizon, for some constant $\omega$ there is a compact spacelike codimension-2 surface $\mathcal{B}$ in which
\begin{equation}
\xi \onb = \lambda + \omega\,\eta \onb = 0 ~,
\end{equation}
$\xi$ being the null Killing vector that generates the Killing horizon. $\mathcal{B}$ is the so-called bifurcation surface, which in our case is just a particular one-dimensional curve located at $r=r_+$. All fields must obey regularity conditions at the bifurcation surface \cite{Wald1993,JacobsonKangMyers}. Following the arguments of \cite{HorowitzWelch-di}, the particular choice of coordinates in a solution like (\ref{BTZsolution}), as well as the gauge choice for the $B$-field, must be such that
\begin{equation}
\mathrm{G}_{t\psi}\big|_{\mathcal{N}} = B_{t\psi}\big|_{\mathcal{N}} = 0 
\end{equation}
at the horizon $\mathcal{N}$. Otherwise, either the original solution or its T-dual along $\psi$ are singular at leading order in $\gamma_\pm$. $\mathrm{G}_{t\psi}\hor=0$ can be enforced using the coordinate $\psi = \psi' - \omega t$, so that:
 \begin{equation}  \label{xiandeta}
 \xi = \partial_t ~, \qquad \eta = \partial_{\psi} ~ .
 \end{equation}
In these new coordinates we are guaranteed\footnote{It may seem that as $\xi$ vanishes only on $\mathcal{B}$ and not on the whole horizon $\mathcal{N}$, $\mathrm{G}_{t\psi}=0$ only holds in $\mathcal{B}$. Nevertheless,  if $\mathcal{N}$ is connected with the bifurcation surface along the integral lines of the Killing field $\xi$, then by symmetry and continuity $\mathrm{G}_{t\psi}=0$ holds in every point of $\mathcal{N}$ \cite{HorowitzWelch-di}.} that $\mathrm{G}_{t\psi} \hor = \mathrm{G}_{M N}\xi^M \eta^N\hor = 0$. The T-duality along $\psi$ and the one along $\psi'$ both correspond to the same Killing vector $\eta$. In what follows we will only use the $\psi$ coordinate, which is indeed the same that already appeared in \eqref{BTZsolution}. Thereby, $B_{t\psi}\hor =0$ can be derived from $B_{t\psi} = B_{MN} \xi^M \eta^N$, and the fact that $\xi$ vanishes on the bifurcation surface $\mathcal{B}$.

On top of that, given the fact that $\Omega_{MA}{}^B$ enters explicitly in the action \eqref{Action}, we want to impose regularity on the Lorentz connection components:
 \begin{equation}
 \xi^M \Omega_{MA}{}^B\hor = \Omega_{tA}{}^B\hor = 0 ~.
 \label{OmegaRegular}
 \end{equation}
This follows from $\xi\onb = \partial_t \onb=0$, and the regularity of $\Omega_{A}{}^B $ as 1-forms. The extension of the result from $\mathcal{B}$ to $\mathcal{N}$ is analogous to that of $\mathrm{G}_{t\psi}\hor = B_{t\psi}\hor =0$, explained just above. With the regular condition $\eqref{OmegaRegular}$, the redefinition \eqref{BuscherBMNPhi} that we will need to do in order to perform T-duality at linear order in $\gamma_\pm$ preserves the generating null Killing vector $\xi = \partial_t$, as established for regular field redefinitions in \cite{JacobsonKangMyers}. It turns out that the obvious vierbein choice of our solution \eqref{BTZsolution},
\begin{equation}
e^0 = N dt ~, \qquad e^1 = \frac{dr}{N} ~, \qquad e^2 = r (d\psi + N^\psi dt) ~,
\label{vierbeinBTZ}
\end{equation}
does not fulfill \eqref{OmegaRegular},
\begin{equation}
\Omega_{t0}{}^1\hor = -\kappa \neq 0 ~,
\label{vierbeinBTZ}
\end{equation}
where
\begin{equation}
\kappa = \frac{r_+^2 - r_-^2}{\ell^2 r_+} = 2\pi T_H ~,
\label{kappa}
\end{equation}
and $T_H$ is the Hawking temperature of the BTZ black hole (we are not considering the extremal cases).\footnote{The time independence of the vierbein can introduce a singularity of the Lorentz connection at the horizon. For this fact and its implications on Wald formalism see \cite{JacobsonMohd}.} As $\Omega_{MA}{}^B$ depends on the vierbein choice, we expect that it can be made regular by a Lorentz transformation. This expectation turns out to be true by applying a local Lorentz boost in the radial direction,
\begin{equation}
E^0 = \cosh (\kappa t)\,e^0 + \sinh (\kappa t)\,e^1 ~, \qquad E^1 = \sinh (\kappa t)\,e^0 + \cosh (\kappa t)\,e^1 ~,
\label{regularvierbeinBTZ}
\end{equation}
and $E^2 = e^2$, which leads to a regular Lorentz connection, $\Omega_{tA}{}^B \hor = 0$. One may wonder if (\ref{regularvierbeinBTZ}) introduces any measurable time dependence in the solution, given that the action is not Lorentz invariant. In the following we show that this is not the case for the fields $\mathrm{G}_{M N}, B_{M N}, \Phi$. 

Given that our action (\ref{Lalphaprime}) is not local Lorentz invariant, we must preserve $\bar H_{MNR}$ for the new vierbein to be a solution; that is, we have to find the anomalous contribution to $B_{M N}$ that we call $\Delta B_{MN}$. It cannot be directly computed from (\ref{anomalouslorentz}) since the local Lorentz transformation (\ref{regularvierbeinBTZ}) is finite. Nevertheless, if we compute the change in $\csform^{(\pm)}_{MNR}$ entailed by (\ref{regularvierbeinBTZ}), we can easily read that of $H_{MNR}$; {\it i.e.}, $B_{MN}$ up to a gauge choice. This last freedom is limited by the regularity condition, $B_{t\psi}\big|_{\mathcal{N}} = 0$. We obtain $B_{t\psi} \to B_{t\psi} + \Delta B_{t\psi}$ with
\begin{equation}
\Delta B_{t\psi} = 2\kappa \frac{r_+-r}{\ell} \left( \gamma_+ - \frac{\gamma_- r_-}{r} \right) ~,
\label{regularBTZ}
\end{equation}
the metric and the dilaton remaining the same as in (\ref{BTZsolution}). Notice that despite the time dependence of the local Lorentz boost \eqref{regularvierbeinBTZ}, the resulting background is manifestly time independent when written in $t,r,\psi$ coordinates and we are still in a stationary solution. Furthermore, the metric is unchanged, thereby we are still dealing with a (regular) bifurcated Killing horizon at $r=r_+$.

%%%%%%%%%%%%%%%%%%%%%%
\section{The $\gamma_\pm$-corrected T-dual black strings}
%%%%%%%%%%%%%%%%%%%%%%

In this section we apply the $\gamma_\pm$-corrected T-duality rules to the BTZ black hole solution \eqref{BTZsolution}, and afterwards we show that the dual can be rewritten as a black string with $\gamma_\pm$ corrections. The detailed derivation of the corrected T-duality rules will be presented in a separate paper \cite{ESSV2}. Here we simply present the recipe to do it and display the final result. The key point is that it is possible to perform a non-covariant field redefinition that keeps the Buscher rules (\ref{BuscherRules}) unaffected \cite{MarquesNunez}. To this end, let us define the background fields $\tilde{\mathrm{G}}_{MN}$, $\tilde{B}_{MN}$, and $\tilde \Phi$,
\begin{eqnarray}  \label{BuscherMetric}
\tilde{\mathrm{G}}_{MN} &=& \mathrm{G}_{MN} - \frac{1}{4} \sum_{k=\pm} a_k\, \Omega_{MA}^{(k)~B}\, \Omega_{NB}^{(k)~A} ~, \\ [0.5em]
\label{BuscherBMNPhi} \tilde{B}_{MN} &=& B_{MN} ~, \qquad e^{-2 \tilde \Phi} \sqrt{-\tilde{\mathrm{G}}} = e^{-2 \Phi} \sqrt{-\mathrm{G}} ~.
\end{eqnarray}
It can be shown that these fields transform according to the Buscher rules; therefore, they have to be applied and, subsequently, the field redefinition (\ref{BuscherMetric}) and (\ref{BuscherBMNPhi}) must be inverted. In this fashion, a dependence in the $\gamma_\pm$ parameters is inherited by the T-dual solution. If we start from the BTZ black hole solution (\ref{BTZsolution}), we obtain:
\begin{eqnarray}
& & ds^2 = -N^2 dt^2 + \frac{dr^2}{N^2} + e^{-2\sigma} \left( d\chi + N^\chi\,dt \right)^2 ~, \nonumber \\ [0.4em]
& & e^{-2\Phi} = r^2 (1 + \Delta_+) ~, \label{generaldualBTZcomoving} \\ [0.4em]
& & B_{t\chi} = \frac{e}{\ell} \left( 1 - \frac{r_+^2}{r^2} \right) - \frac{2}{\ell r^2} \left( \gamma_+ \frac{J}{\ell} \left( 1 - \frac{M \ell^2}{2 r^2} \right) +  \gamma_- \left( M - \frac{J^2}{2  r^2} \right) \right) ~, \nonumber
\end{eqnarray}
where $N^2$ is the same lapse function as in (\ref{lapseBTZ}), and
\begin{equation}
N^\chi = \frac{r_+^2-r^2}{\ell} \left( 1 + e  \Delta_- \right) ~,
\label{Nchi}
\end{equation}
with the following definitions:
\begin{equation} 
\Delta_\pm = \frac{2}{r^2}(\gamma_\pm M + \gamma_\mp J/\ell) ~, \qquad e = \frac{r_-}{r_+} ~, \qquad e^{-\sigma} = \frac{1 - \Delta_+}{r} ~.
\label{Deltapm}
\end{equation}
This solution represents a $\gamma_\pm$-corrected black string,\footnote{Notice that by expressing the metric as in (\ref{generaldualBTZcomoving}), there are also terms quadratic in $\gamma_\pm$; their presence is only meaningful as a way to express the first-order $\gamma_\pm$-corrected vierbein in a more compact way.} as will be clear once we make a further coordinate transformation; the $\gamma_\pm = 0$ case corresponds to the solution found in \cite{HorowitzWelch-di}. The bifurcated Killing horizon is not regular, though. Regularity must be imposed both in the metric and matter fields in order to have a consistent thermodynamics. It can be achieved either by applying the T-duality transformation to the regular BTZ black hole solution or, as it turns out to be equivalent, by performing the local Lorentz boost (\ref{regularvierbeinBTZ}) to (\ref{generaldualBTZcomoving}). It has to be accompanied by the transformation $B_{t\chi} \to B_{t\chi} + \Delta B_{t\chi}$, with
\begin{equation}
\Delta B_{t\chi} = \frac{\kappa}{\ell r^2} \left( \frac{\ell J}{r} \gamma_+ + 2 r \gamma_- \right) ~.
\end{equation}
To better connect our result with those in the literature, we shall rewrite our solution as a three-dimensional black string with $\gamma_\pm$ corrections. First, we extend the leading order coordinate transformation found in \cite{HorowitzWelch-di} to include $\gamma_\pm$ corrections in the following way:
\begin{eqnarray}
t &=& \ell (r_+^2 - r_-^2)^{-1/2} \left( 1 - 2 \ell^{-2} \left( \gamma_+ - e \gamma_- \right) \right) (T + X) ~, \\ [0.4em]
\chi &=& - (r_+^2 -r_-^2)^{1/2} \left[ \left( 1 + 2 \ell^{-2} \left( \gamma_+ + e \gamma_- \right) \right) X + 4 \ell^{-2} e \gamma_- \,T\,\right] ~,
\label{hattedcoordinates}
\end{eqnarray}
where $r^2 = \ell \rho$, thereby $r^2_\pm = \ell \rho_\pm$. If we now further rewrite everything in terms of the conserved quantities of the leading order solution \cite{HorowitzWelch-di}, $\cM = \rho_+$ and $\cQ = - \sqrt{\rho_+ \rho_-}$,
\begin{eqnarray} \label{3dblackstring}
ds^2 &=& - \left( 1 - \frac{\cM}{\rho} \right) \left[ 1 - \frac{4 \cM}{\ell^2 \rho} \left( \gamma_+ + e \gamma_- \right) \right] dT^2 - \frac{4  \left( 1 - e^2 \right) \gamma_- \cQ}{\ell^2 \rho} \left( 1 + \frac{\cM}{\rho} \right) dT\,dX \nonumber \\ [0.35em]
&& + \left( 1 + \frac{e \cQ}{\rho} \right) \left[ 1 + \frac{4 \cQ}{\ell^2 \rho} \left( \gamma_- + e \gamma_+ \right) \right] dX^2 + \left( 1 - \frac{\cM}{\rho} \right)^{-1} \left( 1 + \frac{e \cQ}{\rho} \right)^{-1} \frac{\ell^2 d\rho^2}{4 \rho^2} \nonumber \\ [0.65em]
e^{-2\Phi} &=& \ell \rho + 2 \ell^{-1} \left[ \left( 1 + e^2 \right) \gamma_+ \cM - 2 \gamma_- \cQ \right] ~, \\ [0.65em]
B_{TX} &=& - e - \frac{\cQ}{\rho} + \frac{4 \cQ}{\ell^2 \rho} \left( \gamma_+ \left( \frac{\cM - e \cQ}{2\rho} - 1\right) - \gamma_- \left( \frac{e + e^{-1}}{2} + \frac{\cQ}{\rho} \right) \right) ~, \nonumber \\ [0.65em]
\Delta B_{TX} &=& 2 \kappa \ell^{-1/2} \left( \gamma_+ \cQ \rho^{-3/2} - \gamma_- \rho^{-1/2} \right) ~, \nonumber
\end{eqnarray}
where we used that $e = - \cQ/\cM$. Notice that the angular velocity at the horizon is non-vanishing in spite of the fact that $G_{TX} = 0$ at infinity. In the case of bosonic string theory, $\gamma_- = 0$, the metric becomes diagonal:
\begin{eqnarray}
ds^2 &=& - \left( 1 - \frac{\cM}{\rho} \right) \left( 1 - \frac{2 \cM}{\rho} \frac{\ell^2_s}{\ell^2} \right) d T^2 + \left( 1 + \frac{e \cQ}{\rho} \right) \left( 1 + \frac{2 e \cQ}{\rho} \frac{\ell^2_s}{\ell^2} \right) d X^2 \nonumber \\ [0.3em]
&& + \left( 1 - \frac{\cM}{\rho} \right)^{-1} \left( 1 + \frac{e \cQ}{\rho} \right)^{-1} \frac{\ell^2 d\rho^2}{4 \rho^2} \nonumber~,  \\ [0.6em]
e^{-2\Phi} &=& \ell \left[ \rho +  \left( 1 + e^2 \right) \cM\, \frac{\ell^2_s}{\ell^2} \right] ~, \\ [0.6em]
B_{T X} &=& - e - \frac{\cQ}{\rho} + \cQ \left( \frac{\cM - e \cQ}{\rho^2} - \frac{2}{\rho} \right) \frac{\ell^2_s}{\ell^2} ~, \qquad \Delta B_{TX} =   \frac{\kappa  \cQ \ell_s^2}{\ell^{1/2} \rho^{3/2}  }     ~, \nonumber 
\label{3dblackstringbosonic}
\end{eqnarray}
where we used $\gamma_+ = \frac{1}{2}\alpha^\prime = \frac{1}{2} \ell_s^2$, $\ell_s$ being the string length. This is the leading (bosonic) stringy correction to the black string found in \cite{HorneHorowitzSteif}:
\begin{equation}
ds^2 = - \left( 1 - \frac{\rho_+}{\rho} \right) d T^2 + \left( 1 - \frac{\rho_-}{\rho} \right) d X^2 + \left( 1 - \frac{\rho_-}{\rho} \right)^{-1} \left( 1 - \frac{\rho_+}{\rho} \right)^{-1} \frac{\ell^2 d\rho^2}{4\rho^2} ~.
\label{3dblackstringHHS}
\end{equation}
We can finally obtain the asymptotic form of the metric:
\begin{equation}
ds^2 = - d T^2 + d X^2 + \frac{\ell^2 d\rho^2}{4 \rho^2} ~,
\label{3dblackstringasymp}
\end{equation}
with a linear dilaton, $e^{-2\Phi} = \ell \rho$, and a pure gauge Kalb-Ramond two-form, $B_{T X} = - e$. The line element is that of flat space and the scalar curvature at leading order in $1/\rho$ reads
\begin{equation}
R = \frac{4}{\ell^2 \rho} \left[ \left( \cM - e \cQ \right) \left( 1 + \frac{4\gamma_+}{\ell^2} \right) - 2 \gamma_- \cQ \right] \propto \frac{1}{\rho} ~,
\label{scalarRasymp}
\end{equation}
which is the expected fall-off of an asymptotically flat metric in three dimensions \cite{HorneHorowitzSteif}.

%%%%%%%%%%%%%%%%%%%%%%
\section{T-duality and black hole thermodynamics}
%%%%%%%%%%%%%%%%%%%%%%

We want to study the thermodynamics of the $\gamma_\pm$-corrected black string configuration; in particular, its temperature and entropy. To that end, we bring our solution \eqref{generaldualBTZcomoving} to Eddington-Finkelstein form in which the metric is regular at the horizon through the change of coordinates:
\begin{equation}
dv = dt + \frac{dr}{N^2} ~, \qquad d\tilde{\chi} = d\chi - \frac{N^{\chi}}{N^2} dr ~,
\label{EFCoordinates}
\end{equation}
and the resulting line element is the following:
\begin{equation}
ds^2 = -N^2 dv^2 + 2 dv dr + e^{-2\sigma} (d\tilde{\chi} + N^{\chi} dv)^2 ~.
\label{EFMetric}
\end{equation}
The metric is well-behaved at $r = r_{+}$, regardless of the condition $N(r_{+})=0$. Notice that $N^{\chi}(r_+) = 0$. The one-form normal to the Killing horizon is $n = dr\big|_{\mathcal{N}}$. The surface gravity, $\kappa$, can be defined from the timelike Killing vector, $\xi$, through $\xi^M \nabla_M \xi^N \big|_{\mathcal{N}} = \kappa\,\xi^N\big|_{\mathcal{N}}$, thereby $d(\xi^2)\rvert_{\mathcal{N}} = - 2 \kappa n$,
\begin{equation}
d(\xi^2)\big|_{\mathcal{N}} = \partial_r\!\left[ - N^2 + e^{-2\sigma} (N^\chi)^2 \right] dr\big|_{\mathcal{N}} = - \partial_r N^2\,n \big|_{\mathcal{N}} = - 2 \frac{r_{+}^2 - r_{-}^2}{\ell^2 r_{+}}\,n \big|_{\mathcal{N}}  ~.
\label{dxisquared}
\end{equation}
where in the second equality we used $N^\chi\rvert_{\mathcal{N}} = 0$, and regularity of $N^\chi$ and $\sigma$ at the horizon. Since the lapse of the solution is invariant under T-duality, the value of $\kappa$, which only depends on $N$ and not on $\sigma$ and $N^\chi$, is invariant as well. T-duality preserves the bifurcated Killing horizon and its temperature,
\begin{equation}
\kappa = \frac{r_{+}^2 - r_{-}^2}{\ell^2 r_{+}} = \kappa_{\rm BTZ} ~.
\label{EFSurfaceGravity}
\end{equation}
This generalizes the result obtained in \cite{HorowitzWelch-di} to our $\gamma_\pm$-corrected BTZ black hole \eqref{BTZsolution} and its T-dual \eqref{generaldualBTZcomoving}.

Let us scrutinize now the properties of the black hole entropy under T-duality. The invariance found in \cite{HorowitzWelch-di} was purely a geometrical issue, given that at leading order the Bekenstein-Hawking area law is at place. Far from that, the black hole entropy in the $\gamma_\pm$-corrected theory has to be computed by methods that exceed the applicability of the Iyer-Wald formula \cite{Wald1993,Iyer:1994ys} because the action is not Lorentz invariant. We must modify Wald's derivation along the lines of \cite{Tachikawa} to deal with the anomalous Lorentz invariance. In addition, since our formalism relies on the use of the vierbein, Lie derivatives have to be substituted by Lie-Lorentz derivatives \cite{JacobsonMohd}. The black hole entropy is finally given by the following expression \cite{ESSV2}:
\begin{equation}
S_{BH} = 4 \pi \int_\Sigma d\zeta\, e^{-2 \Phi}  \sqrt{\mathcal{G}_h} \left[ 1 + 2 \gamma_+ \left(  R^{01}_{\;\;\; 01} - 
\frac 3 4 H^{A01} H_{A01} \right) -2\gamma_- \Omega^{A01} H_{A01} \right] ~,
\label{generalentropy}
\end{equation}
where $\Sigma$ denotes the cross section of the horizon on which we integrate. As we are in three dimensions $\Sigma$ is just a curve, parameterized either by $\zeta = \psi$ in the BTZ solution \eqref{BTZsolution} or by $\zeta = \chi$ in the dual black string solution \eqref{generaldualBTZcomoving}. Recall that we have set $16 \pi G_N = 1$ and $\mathcal{G}_h$ is the determinant of the induced metric on the horizon. The last term in \eqref{generalentropy} is anomalous Lorentz invariant to linear order in $\gamma_\pm$ \cite{Solodukhin:2005ah}.

Plugging either the BTZ black hole or the $\gamma_\pm$-corrected black string, we get the same answer:
\begin{equation}
S_{BH} = 8 \pi^2 r_+ + \frac{32 \pi^2}{\ell^{2}} \left( \gamma_+ r_+ + \gamma_- r_- \right) ~,
\label{entropyevaluated}
\end{equation}
where we have set $\Delta \psi = \Delta \chi = 2 \pi$ at the horizon. It is interesting to see how the expressions in the entropy integrand behave when dimensionally reduced (\ref{KMansatz}):
\begin{eqnarray}
& & e^{-2\Phi} \sqrt{\mathcal{G}_h} = e^{-2\phi} ~, \label{Tdualone} \\ [0.6em]
& &  R^{01}_{\;\;\;\;01} - \frac 3 4 H^{012} H_{012} =  \tilde R^{01}_{\;\;\;\;01} - \frac 3 4 \left( e^{2 \sigma} V^{01} V_{01} + e^{-2 \sigma} W^{01} W_{01} \right) ~, \label{Tdualtwo} \\ [0.45em]
& & H^{201} \Omega_{201} = - \frac{1}{2} W^{01} V_{01} ~, \label{Tdualthree}
\end{eqnarray}
where $\tilde R^{\mu\nu}_{\;\;\;\rho\sigma}$ is the dimensionally reduced Riemann tensor, and $V_{\mu\nu}$ and $W_{\mu\nu}$ are the field strengths\footnote{Indices in (\ref{Tdualtwo}) and (\ref{Tdualthree}) are flat, $\tilde R^{ab}{}_{cd}, V_{ab}$ and $W_{ab}$, and they refer to the basis \eqref{regularvierbeinBTZ}.} corresponding, respectively, to $V_\mu$ and $W_\mu$.  Each of the terms is T-dual invariant by itself. Whereas \eqref{Tdualone} is exactly invariant \cite{ESSV2}, \eqref{Tdualtwo} and \eqref{Tdualthree} are only requested to be invariant under the uncorrected Buscher rules (\ref{Tdualityrules}), given that they are already multiplied by $\gamma_\pm$. Notice that for $\gamma_- \neq 0$ the entropy depends on the inner horizon radius $r_-$. This is standard in theories with broken parity,  such as Topologically Massive Gravity \cite{DeserJackiwTempleton} or Mielke-Baekler's gravity \cite{MielkeBaekler,Santamaria:2011cz}.

Given the asymptotics of the BTZ black hole, one may wonder about the consistence of our results from the point of view of the AdS/CFT correspondence. The entropy of the BTZ solution matches the Cardy formula \cite{Cardy}, both in the integrated and the non-integrated versions.\footnote{We would like to thank Gast\'on Giribet for pointing out the possibility to perform this non-trivial check, and for his abundant crucial inputs on this respect.} The central charges of the dual CFT can be readily computed,
\begin{equation}
c_\pm = \frac{3\ell}{2 G_N} + \frac{6}{\ell G_N} (\gamma_{+} \pm \gamma_{-}) ~,
\label{CentralCharges}
\end{equation}
where we have reintroduced $16\pi G_N$ for the sake of comparison. In the case $\gamma_\pm = 0$ we recover the Brown-Henneaux seminal result \cite{BrownHenneaux},
\begin{equation}
c_\pm = \frac{3\ell}{2 G_N} ~.
\label{BrownHenneaux}
\end{equation}
Notice also that these expressions satisfy the integrated version of the Cardy formula,
\begin{equation}
S_{BH} = \frac{\pi^2\ell}{3} (c_+ T_{+} +  c_- T_{-}) ~,
\label{IntegratedCardy}
\end{equation}
where $T_{+}$ and $T_{-}$ are the geometric orbifold temperatures,
\begin{equation}
T_{\pm} = \frac{r_{+} \pm r_{-}}{2\pi\ell^2} ~.
\end{equation}
This CFT description, being non-symmetric in its left- and right-moving sectors due to the bulk Chern-Simons term, yields a quantization condition for $\gamma_{-}$. To show this, notice that modular invariance imposes $c_- - c_+ = 24k$, $k \in \IZ$ \cite{Witten2007}. Thereby, $\gamma_{-}$ must be quantized in units of $2 \ell G_N$. We need to guarantee that $\gamma_- \sim \ell_\star^2$, which can be achieved for $\ell_{\rm Pl} \ll \ell_\star \ll \ell$, where $\ell_\star = M_\star^{-1}$ and $\ell_{\rm Pl} = G_N$.

For heterotic string theory, the couplings are related as $\gamma_+ = - \gamma_-$, and the entropy correction depends on the combination $r_+ - r_-$. This is somewhat consistent with the absence of corrections to the dual background for the extremal heterotic solution, although a bit stronger.

%%%%%%%%%%%%%%%%%%%%%%
\section{Concluding remarks}
%%%%%%%%%%%%%%%%%%%%%%

We have studied the effect of applying T-duality to a BTZ black hole in the presence of higher-derivative corrections parameterized by $\gamma_\pm \sim M_\star^{-2}$. This two-parameter family of theories includes (but generalizes) the next-to-leading $\alpha^\prime$-corrected string theory actions. We used T-duality as a solution generating technique to obtain a non-trivial higher-curvature asymptotically flat black string configuration, which satisfies the equations of motion to the requested order in $M_\star^{-1}$. We show that both the temperature and the entropy remain invariant for all values of $\gamma_\pm$.

The result is puzzling in the following sense. T-duality is not expected to be a symmetry of theories based on point particles. The theories we have just discussed, for generic $\gamma_\pm$ do not seem to have a sigma model origin. T-duality for them might be an approximate symmetry, in the sense that we are always neglecting quadratic contributions in $\gamma_\pm$. The fact that bifurcate Killing horizons are mapped onto bifurcated Killing horizons with exactly the same surface gravity, generalizing the results obtained by Horowitz and Welch \cite{HorowitzWelch-di}, is noticeable. It can be understood from the fact that $\gamma_\pm$-corrected T-duality amounts to a sequence of field redefinitions with an intermediate transformation given by the uncorrected Buscher rules in between. In order to extend the results of \cite{JacobsonKangMyers} to our case, we needed to rely on a notion of regularity for $\Omega_{MA}{}^B$, much in the spirit of \cite{HorowitzWelch-di}. The preservation of entropy, in turn, is remarkable. One might naively expect that neglecting quadratic contributions in $\gamma_\pm$ may affect a quantity that accounts for the number of degrees of freedom. In the case studied by \cite{HorowitzWelch-di} this statement amounts to area conservation (strictly speaking, it is the area in a conformally related metric, which is T-dual invariant), but in the theories studied in this article the entropy is not given by the horizon area.

Our results suggest that T-duality could be a symmetry of a larger class of theories, which, nonetheless, are far from arbitrary since terms organize as in (\ref{Loneplus}) and (\ref{Loneminus}). Among all these theories, what makes string theory special is probably the fact that it can be resummed to all orders in $\gamma_\pm \sim \alpha^\prime$. In order to have a better understanding, one would need to explore the generalization of these results to $\mathcal{O}(\gamma_\pm^2)$. The action becomes rather intractable, as explicitly shown in \cite{LescanoMarques} for the case $\gamma_+ = \gamma_-$, which was originally considered by Hohm, Siegel and Zwiebach \cite{HSZ}. 

It is interesting to point out that our results are in line with expectations naively arising from a recent study performed in \cite{ArvanitakisBlair}. These authors demonstrated the first law of black hole thermodynamics in the Double Field Theory framework, by applying in that context the covariant phase space approach due to Wald and collaborators \cite{LeeWald,Wald1993,Iyer:1994ys}. The very fact that our two-parametric theory (\ref{Action}) can be (actually, it was originally) formulated in the realm of an $O(D,D)$ invariant Double Field Theory makes it plausible in principle to perform a similar analysis to that of \cite{ArvanitakisBlair} as an alternative avenue to reproduce our results. It would be interesting to pursue this computation.

One may wonder how general is the invariance of $\gamma_\pm$-corrected temperature and entropy under T-duality. This is a natural question because the BTZ black hole is a three-dimensional solution with maximal symmetry. The generalization to black holes in higher dimensions is feasible,\footnote{In \cite{CanoOrtin1}, for instance, $\alpha^\prime$-corrected T-duality was performed on a five-dimensional black hole that arises from a brane configuration of effective heterotic superstring theory. The transformation of string and brane charges was studied, and the macroscopic entropy was found to agree with the microscopic result. For a family of self T-dual solutions in the same theory, see \cite{ChimentoOrtin2}.} and is going to be presented elsewhere \cite{ESSV2}. We shall also explore the uses of the AdS/CFT correspondence to further constrain the $\gamma_\pm$ parameters leading to consistent quantum gravities in three dimensions, as well as to scrutinize the would be chiral point where one of the central charges vanishes, and to study the corresponding emergence of a logarithmic branch.

%%%%%%%%%%%%%%%%%%%%%%
\section*{Acknowledgements}
%%%%%%%%%%%%%%%%%%%%%%

We are deeply grateful to Gast\'on Giribet and Diego Marqu\'es for exceedingly interesting remarks and comments on our work. We would like to thank Oleg Evnin, Benjo Fraser, Eric Lescano and Alexandre Serantes for discussions, and St\'ephane Detournay, Olaf Hohm, Ted Jacobson, Tom\'as Ort\'\i n and Sergey Solodukhin for useful correspondence. The work of J.D.E. is supported by MINECO FPA2014-52218, Xunta de Galicia ED431C 2017/07, FEDER, and the Mar\'\i a de Maeztu Unit of Excellence MDM-2016-0692. He wishes to thank Pontificia Universidad Cat\'olica de Valpara\'\i so and Universidad Adolfo Ib\'a\~nez for hospitality, during the visit funded by CONICYT MEC 80150093. He is also thankful to the Physics Department of the University of Buenos Aires, where part of this work was done under the support of the Milstein program. J.A.S.-G. acknowledges support from CUAASC grant of Chulalongkorn University and Spanish FPI fellowship from FEDER grant FPA-2011-22594. A.V.L. is supported by the Spanish MECD fellowship FPU16/06675.

%%%%%%%%%%%%%%%%%%%%%%
%%%%%%%%%%%%%%%%%%%%%%
%%%%%%%%%%%%%%%%%%%%%%


\begin{thebibliography}{99}   
%%%%%%%%%%%%%%%%%%%%%%
%%%%%%%%%%%%%%%%%%%%%%
%%%%%%%%%%%%%%%%%%%%%%

\bibitem{CEMZ}
X.~O.~Camanho, J.~D.~Edelstein, J.~Maldacena and A.~Zhiboedov,
{\it Causality constraints on corrections to the graviton three-point coupling},
JHEP {\bf 1602}, 020 (2016) [arXiv:1407.5597 [hep-th]].
%%CITATION = doi:10.1007/JHEP02(2016)020;%%

\bibitem{DAppollonio} 
G.~D'Appollonio, P.~Di Vecchia, R.~Russo and G.~Veneziano,
{\it Regge behavior saves String Theory from causality violations},
JHEP {\bf 1505}, 144 (2015) [arXiv:1502.01254 [hep-th]].
%%CITATION = doi:10.1007/JHEP05(2015)144;%%

\bibitem{Brandenberger:1988aj} 
R.~H.~Brandenberger and C.~Vafa,
{\it Superstrings in the early universe},
Nucl.\ Phys.\ B {\bf 316}, 391 (1989).
%%CITATION = doi:10.1016/0550-3213(89)90037-0;%%

\bibitem{Veneziano:1991ek} 
G.~Veneziano,
{\it Scale factor duality for classical and quantum strings,}
Phys.\ Lett.\ B {\bf 265}, 287 (1991).
%%CITATION = doi:10.1016/0370-2693(91)90055-U;%%

\bibitem{Gasperini:1992em} 
M.~Gasperini and G.~Veneziano,
{\it Pre-Big Bang in string cosmology,}
Astropart.\ Phys.\ {\bf 1}, 317 (1993) [hep-th/9211021].
%%CITATION = doi:10.1016/0927-6505(93)90017-8;%%

\bibitem{AldayMaldacena} 
L.~F.~Alday and J.~M.~Maldacena,
{\it Gluon scattering amplitudes at strong coupling,}
JHEP {\bf 0706}, 064 (2007) [arXiv:0705.0303 [hep-th]].
%%CITATION = doi:10.1088/1126-6708/2007/06/064;%%

\bibitem{Hull:2009mi}
C.~Hull and B.~Zwiebach,
{\it Double Field Theory,}
JHEP {\bf 0909}, 099 (2009) [arXiv:0904.4664 [hep-th]].
%%CITATION = doi:10.1088/1126-6708/2009/09/099;%%

\bibitem{Hull:2009zb} 
C.~Hull and B.~Zwiebach,
\textit{\it The gauge algebra of Double Field Theory and Courant brackets,}
JHEP {\bf 0909}, 090 (2009) [arXiv:0908.1792 [hep-th]].
%%CITATION = doi:10.1088/1126-6708/2009/09/090;%%

\bibitem{Hohm:2010jy} 
O.~Hohm, C.~Hull and B.~Zwiebach,
{\it Background independent action for Double Field Theory,}
JHEP {\bf 1007}, 016 (2010) [arXiv:1003.5027 [hep-th]].
%%CITATION = doi:10.1007/JHEP07(2010)016;%%

\bibitem{Hohm:2010pp} 
O.~Hohm, C.~Hull and B.~Zwiebach,
{\it Generalized metric formulation of Double Field Theory,}
JHEP {\bf 1008}, 008 (2010) [arXiv:1006.4823 [hep-th]].
%%CITATION = doi:10.1007/JHEP08(2010)008;%%

\bibitem{Geissbuhler:2013uka} 
D.~Geissbuhler, D.~Marqu\'es, C.~N\'u\~nez and V.~Penas,
{\it Exploring Double Field Theory,}
JHEP {\bf 1306}, 101 (2013) [arXiv:1304.1472 [hep-th]].
%%CITATION = doi:10.1007/JHEP06(2013)101;%%

\bibitem{Aldazabal:2013sca} 
G.~Aldazabal, D.~Marqu\'es and C.~N\'u\~nez,
{\it Double Field Theory: a pedagogical review},
Class.\ Quant.\ Grav.\ {\bf 30}, 163001 (2013) [arXiv:1305.1907 [hep-th]].
%%CITATION = doi:10.1088/0264-9381/30/16/163001;%%

\bibitem{MarquesNunez} 
D.~Marqu\'es and C.~A.~N\'u\~nez,
{\it T-duality and $\alpha^\prime$-corrections},
JHEP {\bf 1510}, 084 (2015) [arXiv:1507.00652 [hep-th]].
%%CITATION = ARXIV:1507.00652;%%

\bibitem{Hohm2014} 
O.~Hohm and B.~Zwiebach,
{\it Green-Schwarz mechanism and $\alpha^\prime$-deformed Courant brackets},
JHEP {\bf 1501}, 012 (2015) [arXiv:1407.0708 [hep-th]].
%%CITATION = doi:10.1007/JHEP01(2015)012;%%

\bibitem{Hohm2015} 
O.~Hohm and B.~Zwiebach,
{\it T-duality constraints on higher derivatives revisited},
JHEP {\bf 1604}, 101 (2016) [arXiv:1510.00005 [hep-th]].
%%CITATION = doi:10.1007/JHEP04(2016)101;%%

\bibitem{Buscher:1987sk}
T.~H.~Buscher,
{\it A symmetry of the string background field equations},
Phys.\ Lett.\ B {\bf 194}, 59 (1987).
%%CITATION = doi:10.1016/0370-2693(87)90769-6;%%

\bibitem{KaloperMeissner} 
N.~Kaloper and K.~A.~Meissner,
{\it Duality beyond the first loop},
Phys.\ Rev.\ D {\bf 56}, 7940 (1997) [hep-th/9705193].
%%CITATION = doi:10.1103/PhysRevD.56.7940;%%

\bibitem{MaharanaSchwarz}
J.~Maharana and J.~H.~Schwarz,
{\it Noncompact symmetries in string theory},
Nucl.\ Phys.\ B {\bf 390}, 3 (1993) [hep-th/9207016].
%%CITATION = doi:10.1016/0550-3213(93)90387-5;%%

\bibitem{MetsaevTseytlin}
R.~R.~Metsaev and A.~A.~Tseytlin,
{\it Order $\alpha^\prime$ (two Loop) equivalence of the string equations of motion and the sigma model Weyl invariance conditions: Dependence on the dilaton and the antisymmetric tensor},
Nucl.\ Phys.\ B {\bf 293}, 385 (1987).
%%CITATION = doi:10.1016/0550-3213(87)90077-0;%%

\bibitem{HSZ} 
O.~Hohm, W.~Siegel and B.~Zwiebach,
{\it Doubled $\alpha'$-geometry},
JHEP {\bf 1402}, 065 (2014) [arXiv:1306.2970 [hep-th]].
%%CITATION = doi:10.1007/JHEP02(2014)065;%%

\bibitem{BTZ} 
M.~Ba\~nados, C.~Teitelboim and J.~Zanelli,
{\it The black hole in three-dimensional space-time},
Phys.\ Rev.\ Lett.\ {\bf 69}, 1849 (1992) [hep-th/9204099].
%%CITATION = doi:10.1103/PhysRevLett.69.1849;%%

\bibitem{HorowitzWelch-di} 
G.~T.~Horowitz and D.~L.~Welch,
{\it Duality invariance of the Hawking temperature and entropy},
Phys.\ Rev.\ D {\bf 49}, 590 (1994) [hep-th/9308077].
%%CITATION = doi:10.1103/PhysRevD.49.590;%%

\bibitem{BergshoeffdeRoo} 
E.~A.~Bergshoeff and M.~de Roo,
{\it The quartic effective action of the heterotic string and supersymmetry},
Nucl.\ Phys.\ B {\bf 328}, 439 (1989).
%%CITATION = doi:10.1016/0550-3213(89)90336-2;%%

\bibitem{HorowitzWelch-bh}
G.~T.~Horowitz and D.~L.~Welch,
\textit{Exact three-dimensional black holes in string theory,}
Phys.\ Rev.\ Lett.\ {\bf 71}, 328 (1993) [hep-th/9302126].
%%CITATION = doi:10.1103/PhysRevLett.71.328;%%

\bibitem{HorneHorowitzSteif}
J.~H.~Horne, G.~T.~Horowitz and A.~R.~Steif,
\textit{An Equivalence between momentum and charge in string theory},
Phys.\ Rev.\ Lett.\ {\bf 68}, 568 (1992) [hep-th/9110065].
%%CITATION = doi:10.1103/PhysRevLett.68.568;%%

\bibitem{Hohm2016a}
O.~Hohm,
{\it Background independence and duality invariance in String Theory},
Phys.\ Rev.\ Lett.\  {\bf 118}, 131601 (2017) [arXiv:1612.03966 [hep-th]].
%%CITATION = doi:10.1103/PhysRevLett.118.131601;%%

\bibitem{Hohm2016b}
O.~Hohm,
{\it Background independent Double Field Theory at order $\alpha^\prime$: Metric vs. Frame-like geometry},
Phys.\ Rev.\ D {\bf 95}, 066018 (2017) [arXiv:1612.06453 [hep-th]].
%%CITATION = doi:10.1103/PhysRevD.95.066018;%%

\bibitem{RaczWald} 
I.~Racz and R.~M.~Wald,
\textit{Extension of space-times with Killing horizon},
Class.\ Quant.\ Grav.\ {\bf 9}, 2643 (1992).
%%CITATION = doi:10.1088/0264-9381/9/12/008;%%

\bibitem{Wald1993} 
R.~M.~Wald,
\textit{Black hole entropy is the Noether charge},
Phys.\ Rev.\ D {\bf 48}, R3427 (1993) [gr-qc/9307038].
%%CITATION = doi:10.1103/PhysRevD.48.R3427;%%

\bibitem{JacobsonKangMyers} 
T.~Jacobson, G.~Kang and R.~C.~Myers,
\textit{On black hole entropy},
Phys.\ Rev.\ D {\bf 49}, 6587 (1994) [gr-qc/9312023].
%%CITATION = doi:10.1103/PhysRevD.49.6587;%%

\bibitem{JacobsonMohd} 
T.~Jacobson and A.~Mohd,
\textit{Black hole entropy and Lorentz-diffeomorphism Noether charge},
Phys.\ Rev.\ D {\bf 92}, 124010 (2015) [arXiv:1507.01054 [gr-qc]].
%%CITATION = doi:10.1103/PhysRevD.92.124010;%%

\bibitem{ESSV2}
J.~D.~Edelstein, K.~Sfetsos, J.~A.~Sierra-Garcia and A.~Vilar L\'opez,
{\it to appear}.

\bibitem{Iyer:1994ys} 
V.~Iyer and R.~M.~Wald,
\textit{Some properties of Noether charge and a proposal for dynamical black hole entropy},
Phys.\ Rev.\ D {\bf 50}, 846 (1994) [gr-qc/9403028].
%%CITATION = doi:10.1103/PhysRevD.50.846;%%

\bibitem{Tachikawa} 
Y.~Tachikawa,
\textit{``Black hole entropy in the presence of Chern-Simons terms,''}
Class.\ Quant.\ Grav.\  {\bf 24}, 737 (2007) [hep-th/0611141].
%%CITATION = doi:10.1088/0264-9381/24/3/014;%%

\bibitem{Solodukhin:2005ah} 
S.~N.~Solodukhin,
\textit{Holography with gravitational Chern-Simons},
Phys.\ Rev.\ D {\bf 74}, 024015 (2006) [hep-th/0509148].
%%CITATION = doi:10.1103/PhysRevD.74.024015;%%

\bibitem{DeserJackiwTempleton} 
S.~Deser, R.~Jackiw and S.~Templeton,
\textit{Topologically massive gauge theories},
Annals Phys.\ {\bf 140}, 372 (1982) [Annals Phys.\ {\bf 281}, 409 (2000)] Erratum: [Annals Phys.\ {\bf 185}, 406 (1988)].
%%CITATION = doi:10.1006/aphy.2000.6013, 10.1016/0003-4916(82)90164-6;%%

\bibitem{MielkeBaekler} 
E.~W.~Mielke and P.~Baekler,
\textit{``Topological gauge model of gravity with torsion,''}
Phys.\ Lett.\ A {\bf 156}, 399 (1991).
%%CITATION = doi:10.1016/0375-9601(91)90715-K;%%

\bibitem{Santamaria:2011cz} 
R.~C.~Santamaria, J.~D.~Edelstein, A.~Garbarz and G.~E.~Giribet,
\textit{On the addition of torsion to chiral gravity},
Phys.\ Rev.\ D {\bf 83}, 124032 (2011) [arXiv:1102.4649 [hep-th]].
%%CITATION = doi:10.1103/PhysRevD.83.124032;%%

\bibitem{Cardy} 
J.~L.~Cardy,
\textit{Operator content of two-dimensional conformally invariant theories},
Nucl.\ Phys.\ B {\bf 270}, 186 (1986).
%%CITATION = doi:10.1016/0550-3213(86)90552-3;%%

\bibitem{BrownHenneaux} 
J.~D.~Brown and M.~Henneaux,
\textit{Central charges in the canonical realization of asymptotic symmetries: an example from three-dimensional gravity},
Commun.\ Math.\ Phys.\  {\bf 104}, 207 (1986).
%%CITATION = doi:10.1007/BF01211590;%%

\bibitem{Witten2007} 
E.~Witten,
\textit{Three-dimensional gravity revisited},
arXiv:0706.3359 [hep-th].
%%CITATION = ARXIV:0706.3359;%%

\bibitem{LescanoMarques} 
E.~Lescano and D.~Marqu\'es,
{\it Second order higher-derivative corrections in Double Field Theory},
JHEP {\bf 1706}, 104 (2017) [arXiv:1611.05031 [hep-th]].
%%CITATION = ARXIV:1611.05031;%%

\bibitem{ArvanitakisBlair}
A.~S.~Arvanitakis and C.~D.~A.~Blair,
{\it Black hole thermodynamics, stringy dualities and Double Field Theory},
arXiv:1608.04734 [hep-th].
%%CITATION = ARXIV:1608.04734;%%

\bibitem{LeeWald} 
J.~Lee and R.~M.~Wald,
\textit{Local symmetries and constraints},
J.\ Math.\ Phys.\ {\bf 31}, 725 (1990).
%%CITATION = doi:10.1063/1.528801;%%

\bibitem{CanoOrtin1} 
P.~A.~Cano, P.~Meessen, T.~Ortin and P.~F.~Ramirez,
{\it $\alpha^\prime$-corrected black holes in String Theory},
arXiv:1803.01919 [hep-th].
%%CITATION = ARXIV:1803.01919;%%

\bibitem{ChimentoOrtin2} 
S.~Chimento, P.~Meessen, T.~Ortin, P.~F.~Ramirez and A.~Ruiperez,
{\it On a family of $\alpha^\prime$-corrected solutions of the Heterotic Superstring effective action},
arXiv:1803.04463 [hep-th].
%%CITATION = ARXIV:1803.04463;%%
\end{thebibliography}
\end{document}